\documentclass[fleqn,10pt]{wlscirep}
\title{An MRI Atlas of the Human Fetal Brain: Reference and Segmentation Tools for Fetal Brain MRI Analysis}
\author[1,$^{\dagger}$]{Mahdi Bagheri}
\author[2,3,$^{\dagger}$]{Clemente Velasco-Annis}
\author[2]{Jian Wang}
\author[2,4]{Razieh Faghihpirayesh}
\author[5]{Shadab Khan}
\author[6]{Camilo Calixto}
\author[6]{Camilo Jaimes}
\author[7]{Lana Vasung}
\author[2]{Abdelhakim Ouaalam}
\author[2]{Onur Afacan}
\author[2]{Simon K. Warfield}
\author[3,*]{Caitlin K. Rollins}
\author[1,*]{Ali Gholipour}
\affil[1]{University of California Irvine, Department of Radiological Sciences, Irvine, CA 92617, USA}
\affil[2]{Boston Children's Hospital and Harvard Medical School, Department of Radiology, Boston, MA 02115, USA}
\affil[3]{Boston Children's Hospital and Harvard Medical School, Department of Neurology, Boston, MA 02115, USA}
\affil[4]{Northeastern University, Department of Electrical and Computer Engineering, Boston, MA 02115, USA}
\affil[5]{ADIA Lab., Abu Dhabi, United Arab Emirates}
\affil[6]{Massachusetts General Hospital and Harvard Medical School, Department of Radiology, Boston, MA 02114, USA}
\affil[7]{Boston Children's Hospital and Harvard Medical School, Department of Pediatrics, Boston, MA 02115, USA}
\affil[$^{\dagger}$]{These authors contributed equally to this work.}
\affil[*]{Co-last authors, caitlin.rollins@childrens.harvard.edu, ali.gholipour@uci.edu}

\begin{abstract}
Characterizing in-utero brain development is essential for understanding typical and atypical neurodevelopment. Building on prior spatiotemporal fetal brain MRI atlases, we present the CRL-2025 fetal brain atlas, a spatiotemporal (4D) atlas of the developing fetal brain between 21 and 37 gestational weeks. This atlas is constructed from MRI scans of 159 fetuses with typically developing brains using a diffeomorphic deformable registration framework integrated with kernel regression on age. CRL-2025 uniquely includes detailed tissue segmentations, transient white matter compartments, and parcellation into 126 anatomical regions. It offers significantly enhanced anatomical details over the CRL-2017 atlas and is presented along with a re-release of the CRL diffusion MRI atlas featuring newly created tissue segmentation and labels. We release de-identified, processed subject-level fetal MRI datasets used to generate CRL-2025, providing input–output transparency and reproducibility. We also provide FetalSEG, a deep learning–based multiclass segmentation tool to facilitate automatic fetal brain MRI segmentation. The CRL-2025 atlas and its tools enable scalable fetal brain MRI segmentation, analysis, and neurodevelopmental research for the broader community.

\end{abstract}
\begin{document}

\flushbottom
\maketitle

\thispagestyle{empty}

\section*{Background and Summary}


\indent 
Human cognition and behavior are largely rooted in prenatal and early postnatal brain development, a critical and vulnerable stage of maturation~\cite{rice2000critical,rees2005fetal,o2017fetal,van2020prenatal,lautarescu2020prenatal,wu2024brain,volpe2001perinatal,volpe2022fetal}. During this time, the brain undergoes a dynamic series of neurodevelopmental processes, including neuronal proliferation, neuronal migration, axonal development and retraction (or reorganization), synaptogenesis and apoptosis, the onset of myelination, and the formation of transient developmental zones within the telencephalic wall~\cite{sidman1973neuronal,rakic1988specification,kostovic2002laminar,radovs2006vitro,kostovic2007transient,bystron2008development,kostovic2009insights,ball2013development,kostovic2014perinatal,vasung2016quantitative}. These compartments provide the scaffolding that lays the foundation for the permanent cortical architecture seen in the adult brain. As a result, accurately characterizing early brain development in utero is essential to understand typical neurodevelopment and identify deviations from normal development. Fetal brain Magnetic Resonance Imaging (MRI), including quantitative MRI analysis enables the study of these processes \textit{in vivo} , but requires the incorporation of accurate brain atlases as templates for spatial normalization and as references defining normative spatiotemporal developmental trajectories. Consequently, significant efforts have been made to build early brain development atlases~\cite{habas2010spatiotemporal,serag2012construction,dittrich2014spatio,gholipour2017normative,schuh2018unbiased,khan2019fetal,wu2021age,calixto2025detailed} using MRI, a modality that provides unmatched images of brain anatomy throughout the human lifespan including, in specific, the fetal period~\cite{bethlehem2022brain,calixto2024advances}. 

\indent Constructing atlases from fetal MRI presents distinct challenges compared to adult atlas generation, primarily due to two factors. First, there are technical challenges in scanning fetuses with MRI. MRI is very sensitive to motion, and fetuses move near continuously during MRI acquisitions~\cite{vasung2023cross}. Intermittent fetal and maternal respiratory movements prohibit precise spatiotemporal encoding that is necessary for real, high-resolution 3D MRI of the fetal brain anatomy. This challenge is exacerbated by the low level of MRI signal that is received from MRI body coils from the small fetal anatomy. 
Despite these challenges, the landscape of fetal MRI has changed dramatically in the past decade by the significant advances in retrospective slice-to-volume reconstruction (SVR) techniques~\cite{Rousseau2006, Jiang2007,gholipour2010robust, kim2010intersection, kuklisova2012reconstruction, tourbier2015efficient, Kainz2015fast,tourbier2017automated,ebner2020automated, uus2020deformable,uus2023retrospective,xu2023nesvor} and the subsequent development of fetal brain MRI atlases and processing tools~\cite{habas2010spatiotemporal,serag2012construction,dittrich2014spatio,gholipour2017normative,schuh2018unbiased,khan2019fetal,wu2021age,calixto2025detailed}. 
Second, the brain structure and function change rapidly and dramatically during the fetal period of development. As a result, single static or 3D atlases are insufficient to capture these dynamic processes. Consequently, atlases covering this period should be spatiotemporal (i.e., dynamic or 4D) to accurately reflect the continuous and age-dependent nature of fetal brain development.

The construction of digital spatiotemporal MRI atlases of early brain development is relatively new: Kuklisova-Murgasova et al. \cite{kuklisova2011dynamic} developed a 4D probabilistic atlas of early brain growth from \textit{in-vivo} MRI of 142 preterm infants in the 29 to 44 weeks post-menstrual age. They used pairwise affine registration of anatomy with kernel regression in age for atlas construction. Serag et al.\cite{serag2012construction} used a non-rigid registration approach based on Bspline free-form deformations (FFD) \cite{rueckert1999nonrigid} and showed a marked improvement over the use of affine registration in atlas construction. Makropoulos et al.\cite{makropoulos2016regional} used a similar approach to construct a probabilistic spatiotemporal atlas of the neonatal brain from 420 segmented MRIs of neonates (including preterm neonates) scanned between 27 to 45 weeks post-menstrual age. To improve the FFD-based 4D atlas construction framework, Schuh et al.\cite{schuh2014construction} developed diffeomorphic registration based on the Log-Euclidean mean of inverse consistent FFD transformations.

\indent 
For the fetal period of brain development, a recent review~\cite{ciceri2024fetal} found 18 atlases presented in the literature, of which 12 are publicly available. Most of these atlases characterize the anatomy (structure) of the fetal brain based on 3D-reconstructed T2-weighted (T2w) MRI, but a few atlases also characterize the microstructure of the fetal brain based on diffusion-weighted MRI (dMRI) or the cortical surface based on cortical surface meshes reconstructed from anatomical MRI.

The first fetal MRI atlas was developed by Habas et al.\cite{habas2010spatiotemporal}, based on 20 normal-appearing fetal brains scanned between 21–24 gestational weeks (GAs), offering age-specific T2w MRI templates and tissue probability maps of key brain structures such as cortical gray matter (cGM), white matter (WM), germinal matrix, and lateral ventricles (LV), using manual segmentation followed by groupwise registration and polynomial modeling of structural changes. Serag et al.\cite{serag2012construction} used non-rigid FFD-based registration along with adaptive kernel regression on age to build a spatiotemporal atlas of fetal brain anatomy in a much broader age range of 23 to 37 gestational weeks. This atlas was built by using 3D-reconstructed~\cite{kuklisova2012reconstruction} T2w brain MRI images of 80 typically-developing fetuses. Dittrich et al.\cite{dittrich2014spatio} created an atlas from 32 fetuses between 20–30 GA using a semi-supervised method to reduce manual annotation by segmenting from partially labeled data. Gholipour et al.\cite{gholipour2017normative} built a spatiotemporal atlas of the fetal brain anatomy from super-resolution reconstructed~\cite{gholipour2010robust} fetal brain MRIs of 81 healthy fetuses, covering 21-38 GA. This atlas, referred to as the CRL (Computational Radiology Lab) Fetal Brain Atlas, was built using kernel regression on age integrated into an ANTS~\cite{Avants2008} group-wise symmetric diffeomorphic deformable registration framework. The CRL atlas uniquely offers detailed labels and tissue segmentations, and has been used as the standard space in several fetal SVR tools such as NiftyMIC~\cite{ebner2020automated}, MIALSRTK~\cite{tourbier2017automated}, and NeSVoR~\cite{xu2023nesvor}, and was used as a reference for comparison to the fetal brain ultrasound atlas developed by Namburete et al.~\cite{namburete2023normative}. 

Urru et al.\cite{urru2023automatic} aligned the CRL fetal atlases with neonatal atlases from Serag et al.\cite{serag2012construction}, to build a unified fetal and neonatal label representation to support perinatal brain segmentation and analyses. In other atlas construction efforts, Li et al.~\cite{li2015construction} constructed a fetal brain atlas from 35 Chinese subjects between 23–36 GAs at 2-week intervals, using deformable registration without explicit region of interest (ROI) definitions. Wu et al.~\cite{wu2021age} enhanced this by generating weekly templates from 89 Chinese fetuses between 21–35 GAs using MIRTK registration and label propagation from the CRL atlas, making it the first publicly available atlas focused on this population. Xu \& Sun et al.\cite{xu2023nesvor} built upon these efforts by using 90 high-resolution fetal MRIs acquired on a 3T scanner from 23–38 GAs to construct an atlas with 85 brain structures, including the hippocampus and amygdala, segmented using the Draw-EM tool. Fidon et al.\cite{fidon2024dempster} presented the first fetal atlas targeting a clinical population—Spina Bifida Aperta (SBA)—from 37 fetuses between 21–34 GAs, using anatomical landmarks, nonlinear registration, and post-processed segmentations to create templates with labels for structures such as the corpus callosum, cerebellum, brainstem, and extra-axial CSF, supporting surgical planning and training of disease-specific deep learning models.

Khan et al.~\cite{khan2019fetal} developed the first spatiotemporal atlas of the fetal brain microstructure based on fetal dMRI. This atlas was constructed from motion-robust reconstructed diffusion tensor images (DTI)~\cite{marami2017temporal} of 67 fetuses, covered the GA range of 23 to 38 weeks, and was released and used along with the CRL T2w atlas to characterize fetal brain maturation in several studies, e.g.~\cite{mallela2023heterogeneous,calixto2024characterizing,calixto2025detailed}. Notably, based on this atlas and the fetal DTI data, Calixto et al.~\cite{calixto2025detailed} have recently developed a spatiotemporal atlas of 60 distinct white matter tracts, including commissural, projection, and association fibers of the fetal brain between 23 and 35 weeks of GA (included in the CRL-2025) Atlas. In other works, Chen et al.~\cite{chen2022deciphering} developed a DTI atlas of the fetal brain from dMRI scans of 89 fetuses from the Chinese population scanned in the GA range of 24 to 38 weeks. 

Uus et al.~\cite{uus2023multi} have built a high-quality, high-resolution multi-modal spatiotemporal MRI atlas of the fetal brain in the GA range of 21 to 36 weeks from reconstructed T2w, T1-weighted (T1w), and dMRI scans of 187 fetuses from the developing Human Connectome Project (dHCP)~\cite{price2019developing}. This atlas, referred to as the dHCP fetal MRI atlas, contains T1w, T2w, and dMRI modalities, including fractional anisotropy, mean diffusivity, and orientation distribution function images of the fetal brain, as well as tissue segmentations including the developing white matter, cortical gray matter, cerebrospinal fluid, and major deep gray matter structures, with a total of 17 labels.  

\indent Our goal in this work was to leverage the most recent technical advances in fetal MRI processing along with improved fetal imaging on 3T MRI scanners to generate a spatiotemporal MRI atlas of the fetal brain with much higher quality than the 2017 CRL atlas, but also with unique and detailed segmentations of the fetal brain anatomy, including tissue segmentations with the transient compartments of the developing white matter (31 total structures), as well as anatomical parcellation (126 total structures) that do not currently exist in any other fetal MRI atlas. These detailed segmentations and labels are crucial to conduct studies on regional brain maturation, connectivity analysis, regional groupwise analysis, and gene expression analysis. Therefore, the CRL-2025 atlas described here fills a critical gap in the field by providing detailed anatomical parcellations and transient compartments for the fetal brain on an anatomically-refined spatiotemporal MRI atlas. In this paper, we present the CRL-2025 fetal brain atlas, its construction framework, atlas labels, and, importantly, an open source framework based on state-of-the-art registration, parcellation, and segmentation methods to support automatic segmentation and population-level neurodevelopmental analysis. 
The CRL-2025 spatiotemporal fetal brain atlas provides a high-resolution anatomical reference from 21 to 37 weeks of GA based on carefully processed scans of 159 fetuses, and is released along with the CRL dMRI atlas~\cite{khan2019fetal} as well as tissue segmentations and regional parcellations on the dMRI atlas, which are also novel to the field. The dMRI atlas provides diffusion tensor images and derived scalar maps (fractional anisotropy (FA), color FA (CFA) and mean diffusivity) as well as detailed tissue segmentations and parcellations. 

Together or separately, these atlases can be used in a wide range of applications such as spatial normalization, automatic segmentation through label propagation, regional morphometric and volumetric analysis, multi-modal analysis, tract segmentation, tractography, and connectivity analysis to study developmental trajectories at the population level. To enable rapid and accurate region-level analysis on both T2w and dMRI scans, we have developed and released segmentation and parcellation pipelines based on (1) multi-atlas segmentation and (2) deep learning - two complementary approaches that balance anatomical fidelity with computational scalability. 
To facilitate automatic fetal brain MRI segmentation on reconstructed fetal MRI scans, we trained and evaluated several deep learning models based on state-of-the-art network architectures, including nnU-Net~\cite{isensee2021nnu}, UNETR~\cite{hatamizadeh2021unetr}, Swin UNETR~\cite{hatamizadeh2021swin,cao2022swin}, and an attention-guided lightweight Mamba network~\cite{liao2024lightmunet}. 
Among the evaluated models, the model based on nnU-Net demonstrated the best overall performance and robustness; therefore, we report only the results of this model and provide the trained model as a practical tool, named FetalSEG, for fetal brain MRI segmentation. These tools and atlases have been publicly released to support the research community in advancing automated fetal brain MRI analysis.

\section*{Methods}
\label{sec_methods}

\subsection*{Fetal MRI data}
\indent Fetal MRI data used for the construction of the CRL-2025 Fetal T2 atlas were acquired on several MRI scanners through research studies conducted at Boston Children's Hospital between 2014 and 2023. All studies and the imaging protocols were reviewed and approved by the Institutional Review Board (IRB- P00002874, P00019668, P00034405, P00041613, P00041916, P00008836), and written informed consent was obtained from all participants. Inclusion criteria for those studies were for pregnant women between the ages of 18 and 45 who volunteered for research fetal MRIs and did not have any contraindication for MRI. Exclusion criteria for this study included anyone with a contraindication for MRI, multiple gestation pregnancies, or fetuses with any known, suspected, or documented morphological brain abnormality or diagnosis.
Scans that were eventually included in the atlas construction were selected from a large database based on both acquisition and reconstruction quality, prioritizing high signal-to-noise ratio and quality of the reconstructed fetal brain images. Following these criteria, a total of 193 fetal MRIs obtained from 159  fetuses were used for spatiotemporal atlas construction. Individuals were scanned up to three times, generally at least six weeks apart. For the majority (129) of the 159 participants only a single scan was used for atlas construction, while 28 participants contributed images from two separate visits, and 3 participants contributed images from three visits.

\indent Fetal MRIs were performed on Siemens 3T MRI scanners (Siemens Healthineers, Erlangen, Germany): Trio (n=5),  Skyra (n=134), Prisma (n=53), or Vida (n=1), and a single scan on a Philips 1.5T scanner. Generally, 18-channel body matrix coils were used until August 2017, after which a 30-channel body matrix coil was used. Full research fetal MRI sessions took up to 60 minutes (with up to 45 minutes of imaging) during which several image sequences were taken based on project needs at the time, with 10-20 minutes of the session dedicated to T2-weighted Half-Fourier Acquisition Single shot Turbo Spin Echo (HASTE) imaging, about 10-15 minutes to diffusion MRI (dMRI), and the rest to other sequences. Typical imaging parameters for HASTE sequences were: echo time 115–120 ms, repetition time 1400-1600 ms, flip angle between 120 and 160, slice thickness = 2–3 mm, in-plane resolution = 1 mm, which was achieved by using a variable matrix size ($256\times256-320\times320$) that was adjusted based on a variable field-of-view (256-320 $mm^2$) to cover the anatomy (mother and fetus). The dMRI scans comprised of 2–8 scans each along one of the orthogonal planes (axial, coronal, sagittal) with respect to the fetal head. In each scan, 1 or 2 b=0s/mm2 images, and 12 diffusion-sensitized images at b=500s/mm2 were acquired. Acquisition parameters were: minimal TR (typically 3000–4000ms), TE=60ms, in-plane resolution=2mm, slice thickness=2–4mm. 

\begin{figure}[!h]
\centering
\includegraphics[width=12cm]{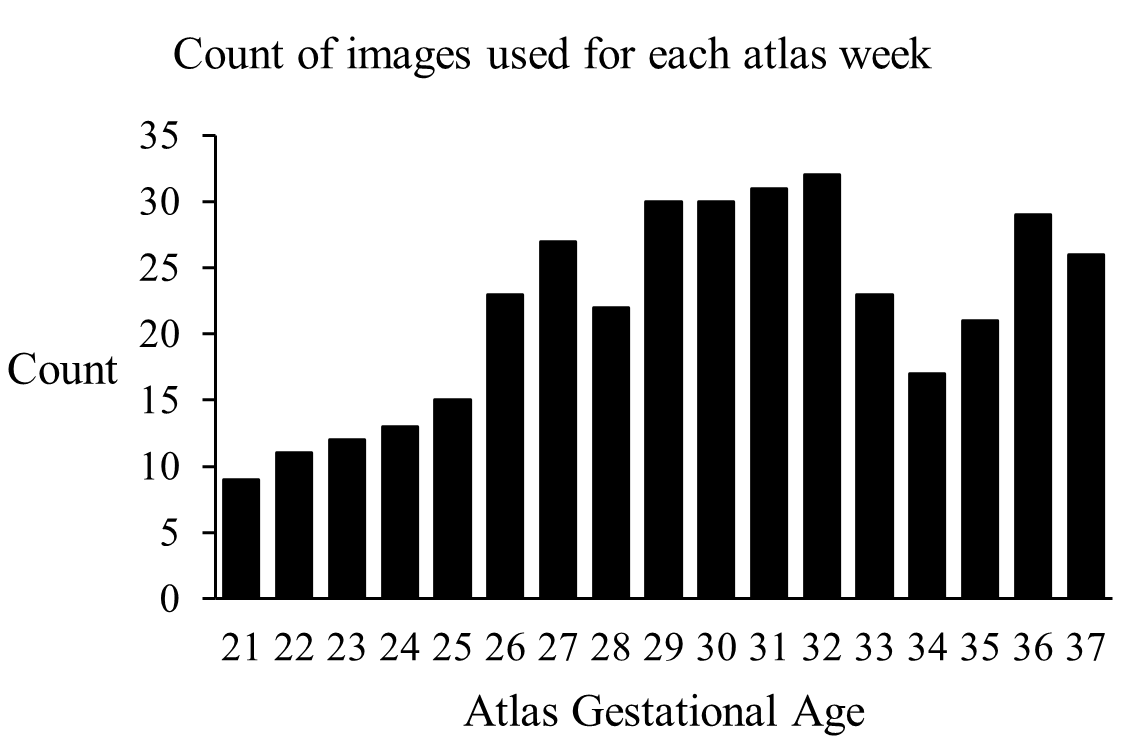}
\caption{Age distribution of samples used in CRL2025 atlas construction. We used Gaussian kernel regression on age with a normalized $\pm$1~week kernel width.}
\label{fig_frequencydist}
\end{figure}

A cutoff window of one gestational week was used to select subjects that contributed to atlas construction at every gestational week; so Fig.\ref{fig_frequencydist} shows the histogram of the gestational age of the subjects used for atlas construction at every atlas week. Fig.\ref{fig_pipeline} shows an overview of the atlas construction process, which is discussed in the sections that follow. It should be noted that while we have built and released atlases at every gestational week, with the presented atlas construction framework, atlases can be built at any continuous age point (i.e., at any fractions of weeks or days).

\subsection*{{Pre-processing of T2w Images}}
\label{sec_preprocessing}

\begin{figure}
\centering
\includegraphics[width=\textwidth]{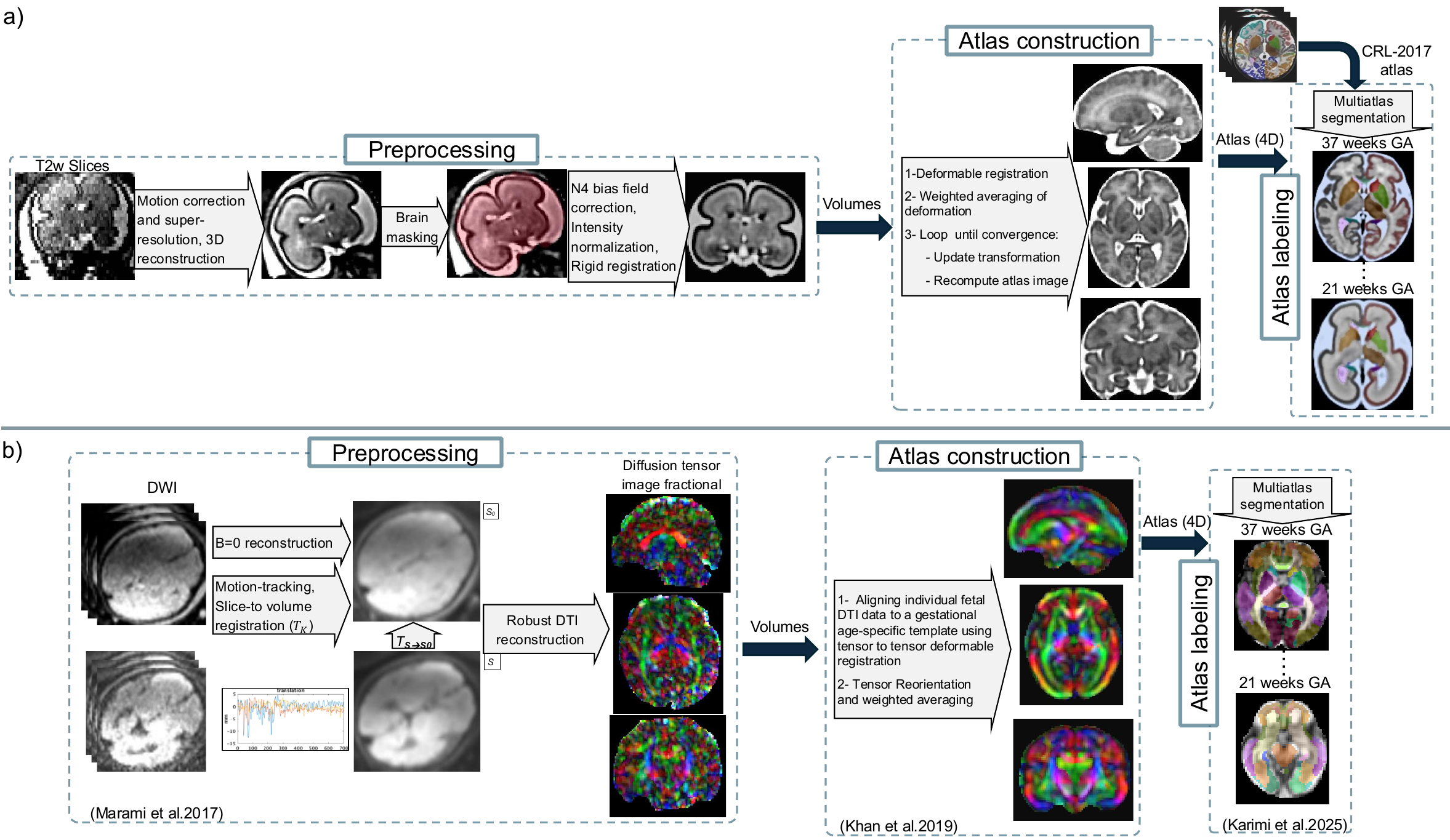}
\caption{Overview of fetal MRI processing for spatiotemporal atlas generation. 
(a) Spatiotemporal anatomical (T2-weighted) atlas generation process based on fast T2-weighted MRI scans. 
(b) Spatiotemporal diffusion MRI atlas generation process based on DTI reconstruction, which is a separate, modality-specific pipeline including 
1) motion-tracking–based slice-to-volume registration for robust diffusion tensor image reconstruction~\cite{marami2017temporal}, 
2) diffusion tensor atlas construction~\cite{khan2019fetal}, and 
3) diffusion tensor atlas labeling~\cite{calixto2025detailed}. 
These two workflows are independent (T2w for structural atlas, DTI for diffusion atlas) rather than alternative preprocessing paths. 
Data, atlas construction, labeling procedures, automatic segmentation methods, and validations are discussed in this article.
}
\label{fig_pipeline}
\end{figure}

\indent
Pre-processing of structural T2-weighted HASTE scans, illustrated in the top row of Fig.\ref{fig_pipeline}, included the following steps: 1. Individual stacks were excluded if severe fetal or maternal motion persisted throughout the stack or image artifacts obscured the fetal brain; 2. Super-resolution reconstruction was performed using either the SVRTK~\cite{uus2020deformable} or NiftyMIC~\cite{ebner2020automated} toolkits.
The choice of reconstruction method was data-driven. NiftyMIC reconstructions were generally chosen due to their superior SNR and tissue border fidelity. In cases where the automated NiftyMIC processing pipeline failed, SVRTK output was used instead if data quality was high. These were therefore not two alternative user workflows, but complementary tools selected pragmatically to ensure optimal reconstruction quality.
3. B0-field inhomogeneity correction with the N4 algorithm~\cite{tustison2010n4itk} and intensity normalization; 4. Intracranial cavity segmentation of the reconstructed brain image ("brain masking"); 5. Rigid registration to atlas anatomical space. Steps 3-4 were performed with in-house tools when SVRTK reconstruction was used, whereas NiftyMIC includes these steps in its reconstruction pipeline. Rigid registration to atlas anatomical space was performed with FLIRT~\cite{jenkinson2012fsl} using six degrees of freedom (translation and rotation only). 
Affine scaling was intentionally not applied in order to preserve the natural size differences across gestational ages, consistent with previous spatiotemporal fetal atlas construction frameworks~\cite{gholipour2017normative,schuh2018unbiased}.

\indent Default settings were used for SVRTK and NiftyMIC was run with the alpha parameter
set to 0.04, chosen for total-variation regularization strength in order to match regularization strength in SVRTK. In our experiments, this value provided stable high-quality
reconstructions across the dataset. Prior to SVRTK reconstruction, a rough ellipsoid mask was drawn over the fetal brain in one reference stack in ITK-SNAP~\cite{yushkevich2006user}. Prior to NiftyMIC reconstruction, the NiftyMIC fetal brain extraction pipeline was used to crop the input images. Each reconstruction included three to eighteen stacks (mean=8 stacks), with at least one axial, one coronal, and one sagittal stacks without significant motion artifacts.

\subsection*{Spatiotemporal anatomical atlas construction}
\label{sec_atlasconstruction}
For spatiotemporal atlas construction, we used a modified recursive mean intensity function to incorporate an estimate of the longitudinal deformations of the atlas space across time points, as presented by Schuh et.al. \cite{schuh2018unbiased}.

To construct the spatiotemporal atlas, we defined the mean intensity and anatomical shape at iteration $k$ as:
\begin{equation}
\bar{I}_k(t) = \sum_{i=1}^{n} w_i(t) \left( \tilde{I}_i \circ {T}_{k,i}^{-1}(t) \right)
\end{equation}
Here, $\tilde{I}_i(x)$ refers to the globally normalized image for subject $i$, and $T_{k,i}$ denotes the estimated diffeomorphic mapping from the subject’s native space to the common atlas space at time $t$, during iteration $k$.

The temporal regression weights $w_i(t)$ are derived from normalized Gaussian functions centered at age $t$, with variance $\sigma_t^2$ adapting to the time point. Specifically:
\( \sigma_t \), i.e., $w_i(t) = \frac{g_i(t)}{\sum_{j=1}^{n} g_j(t)}$
where
$g_i(t) = \frac{1}{\sigma_t \sqrt{2\pi}} \exp\left( \frac{-(t_i - t)^2}{2\sigma_t^2} \right)$.
To restrict the spatial influence of far-off time points, all weights outside of a kernel width of one week were truncated.
The mapping $T_{k,i}(t)$ is composed of a residual deformation $\bar{\phi}_k(t)$ and a subject-specific transformation $\phi_{k,i}$:
\begin{equation}
{T}_{k,i}(t) = {\bar{\phi}}_k(t) \circ \phi_{k,i}
\end{equation}
where
\begin{equation}
\bar{\phi}_k(t) = \exp \left( - \sum_{i=1}^{n} w_i(t) \log \phi_{k,i} \right)
\end{equation}
denotes the weighted Log-Euclidean mean~\cite{arsigny2006log} of the inverse subject-to-atlas deformations, \(\phi_{k,i}\),
obtained via  stationary velocity
free-form deformation (SVFFD) registration of \(\tilde{I}_i\), presented in \cite{schuh2018unbiased}, to the age-matched average image from the previous iteration, i.e., \(\bar{I}_{k-1}(t_i)\). The logarithmic maps \(\log \phi_{k,i}\) correspond directly to the stationary velocity fields (SVFs) computed during registration. The cross-sectional diffeomorphism mapping $\tilde{I}_i$ into the atlas space at $t_i$ is given by
\begin{equation}
\label{eq4}
\phi_{k,i} = \exp(\mathbf{v}_{k,i})
\end{equation}
where
\begin{equation}
\mathbf{v}_{k,i} = \arg\min_{\mathbf{v}} \mathcal{E}(\tilde{I}_i, \bar{I}_{k-1}(t_i), \mathbf{v})
\end{equation}
is the (local) minimum of the energy given presented, and $\mathbf{v}_0 = 0$.
The energy functional $\mathcal{E}(I_0,I_1;\mathbf{v})$ follows the symmetric formulation of Schuh et~al.~\cite{schuh2018unbiased}, combining a local normalized cross-correlation data term with linear-elasticity, bending, and Jacobian-based regularization terms to enforce smooth, topology-preserving deformations.

To track the sequence of spatial deformations applied at each time point \( t \), and to use these deformations to derive a \textit{spatio-temporal coordinate map} (referred to as the longitudinal coordinate map in Schuh et al.~\cite{schuh2018unbiased}) that relates an observed age \( t_i \) to any other time point, thereby correcting for anatomical mismatches in the co-domain, i.e.,
\begin{equation}\label{eq6}
\psi_{k,i}(t) = \bar{\varphi}_k(t) \circ \psi_{k-1,i}(t) \circ \bar{\varphi}_k^{-1}(t_i)
= \left( \prod_{s=1}^{k} \bar{\varphi}_s(t) \right)
\left( \prod_{s=1}^{k} \bar{\varphi}_s(t_i) \right)^{-1}
\end{equation}
where, $\prod$ denotes a sequential composition of functions, and $\bar{\varphi}$ corresponds to the refined Log-Euclidean mean of the transformations. 
The co-domain refers to the spatial coordinate system of the evolving atlas. The longitudinal deformation $\psi_{k,i}(t)$ corrects for anatomical mismatches between different atlas ages by mapping each subject’s anatomy from its acquisition age $t_i$ to a target age $t$, thereby maintaining temporal consistency of anatomical correspondence across gestation. This formulation, which avoids recursion, is based on the assumption that the initial longitudinal deformation between any two time points—specifically from $t_i$ to $t$—is the identity map, i.e., $\psi_{0,i}(t) = \text{Id}$. As each transformation is expressed via the exponential of a SVF, their compositions can be efficiently approximated using the Baker–Campbell–Hausdorff expansion. This ensures that the resulting transformations stay consistent within the SVFFD model.

The composition of the subject-to-atlas transformation given by Equation~(\ref{eq4}), obtained by registering the \(i\)-th image to the template \(\bar{I}_{k-1}(t_i)\), with the longitudinal deformation defined in Equation~(\ref{eq6}), yields an age-specific deformation, i.e.,
\begin{equation}
\varphi_{k,i}(t) = \psi_{k-1,i}(t) \circ \phi_{k,i}
\end{equation}
The residual atlas deformation is now given by the Log-Euclidean mean of the age-adjusted atlas-to-subject transformations. It is thus given by
\begin{equation}
\bar{\varphi}_k(t) = \exp \left( - \sum_{i=1}^{n} w_i(t) \log \varphi_{k,i}(t) \right)
\end{equation}
which are further used to redefine the total age-dependent deformations, i.e.,
\begin{equation}
T_{k,i}(t) = \bar{\varphi}_k(t) \circ \varphi_{k,i}(t)
= \bar{\varphi}_k(t) \circ \psi_{k-1,i}(t) \circ \phi_{k,i}
\end{equation}

\noindent Based on the presented equation, Algorithm~\ref{alg:atlas_construction} is used to construct the spatio-temporal atlas.
{\small
\begin{algorithm}
\caption{Spatio-temporal Atlas Construction}
\label{alg:atlas_construction}
\begin{algorithmic}[1]
\State \textbf{Input:} Globally normalised images $\tilde{I}_i$, previous transformations $\mathbf{T}_{k-1,i}$, weights $w_i(t)$
\State \textbf{Output:} Updated atlas templates and transformations
\ForAll{$t_i$ such that $|t_i| \leq n$}
    \State Generate template images $\bar{I}_{k-1}(t_i)$ given $\tilde{I}_i$ and $\mathbf{T}_{k-1,i}$ \Comment{Eq. (1)}
    \State Compute $\phi_{k,i} = \exp(\mathbf{v}_{k,i})$ from $i$-th image to template at $t_i$ \Comment{Eq. (4)}
    \ForAll{$t \in \{t_j \mid w_i(t_j) > 0\}$}
        \State Compose maps $\psi_{k-1,i}$ given $\psi_{k-2,i}$ and $\bar{\varphi}_{k-1}$ \Comment{Eq. (6)}
        \State Compose maps $\phi_{k,i}$ with $\psi_{k-1,i}$ \Comment{Eq. (7)}
    \EndFor
\EndFor
\ForAll{observed $t_i$}
    \State Compute Log-Euclidean means $\bar{\varphi}_k$ \Comment{Eq. (9)}
\EndFor
\end{algorithmic}
\end{algorithm}
}


\subsection*{Atlas labeling and segmentation}
\label{sec_atlaslabels}

The spatiotemporal fetal brain MRI atlas offers a high signal-to-noise representation of normal fetal brain anatomy across gestation, serving as a robust foundation for tissue-type segmentation and anatomical labeling. Segmentations for the T2-weighted CRL-2025 Atlas were initialized utilizing the pre-existing CRL-2017 atlas as well as reconstructed and labelled individual subject images, as described and validated for image segmentation in Gholipour et al. 2017~\cite{gholipour2017normative} and Rollins et al. 2021~\cite{rollins2021regional}. 

\indent In short, Advanced Normalization Tools (ANTs)~\cite{Avants2008} were first used to perform symmetric diffeomorphic registration of reference atlas images within one week gestational age to each CRL-2025 Atlas image, followed by automatic multi-atlas segmentation with the Probabilistic STAPLE algorithm~\cite{AkhondiAsl2013}, producing two segmentations: 1. a tissue segmentation delineated by structure (including labels such as cortical plate, white matter, and cerebrospinal fluid (CSF) spaces) and 2. a regional segmentation organized by location (gyri, sulci, lobe, etc.), roughly corresponding to the atlas labels presented in Blesa et al. 2016~\cite{blesa2016parcellation}. Full lists of tissue and regional segmentation labels are included in the dataset label keys.

\indent CRL-2025 atlas tissue labels were inspected and manually corrected in ITK-SNAP by an expert with ten years of experience in fetal MRI annotation, following the process detailed in Gholipour et al.~\cite{gholipour2017normative}. White matter compartment labels were only applied to atlas images up to gestational age 32 weeks, as these compartments became less visible with advanced gestational age~\cite{vasung2020quantitative,gholipour2017normative}.
Regional parcellations were generated automatically using multi-atlas probabilistic fusion and assessed for general quality but not systematically refined by hand.

\indent The tissue segmentation protocol has undergone changes compared to the original CRL-2017 atlas~\cite{gholipour2017normative}, which was supplemented with white matter compartment labels in 2020~\cite{rollins2021regional,vasung2020quantitative}. In CRL-2025, regions previously belonging to the hippocampal commissure and subthalamic nuclei have been incorporated into surrounding labels due to low confidence in their accuracy and lack of clinical relevance at this resolution, attributable to their small size and low contrast. New labels were assigned to the CSF-filled spaces of the cavum septum and the third and fourth ventricles, and laterally divided vermis labels were added to the cerebellum. Full lists of each labeling schemes (T2w tissue segmentation, T2w regional segmentation, and diffusion segmentation) are included in the dataset. 

The main differences between the CRL-2017 and CRL-2025 tissue label schemes are summarized in Table~\ref{tab:label_changes_2017_2025}.
\begin{table}[h]
\centering
\caption{Main differences between the CRL-2017 and CRL-2025 tissue label schemes.}
\label{tab:label_changes_2017_2025}
\begin{tabular}{lll}
\hline
CRL-2017 label & CRL-2025 label & Notes on CRL-2025 scheme \\
\hline
CSF & Cavum pellucidum & Subdivision of CRL-2017 CSF label \\
CSF & Third ventricle & Subdivision of CRL-2017 CSF label \\
CSF & Fourth ventricle & Subdivision of CRL-2017 CSF label \\
Cerebellum (left/right) & Vermis & Subdivision of CRL-2017 cerebellum label \\
Subthalamic nuclei (left/right) & Thalamus and white matter (left/right) & Assigned to neighboring regions as appropriate \\
Hippocampal commissure & CSF and corpus callosum & Assigned to neighboring regions as appropriate \\
\hline
\end{tabular}
\end{table}

\indent In total, the tissue segmentation includes 31 labels and the regional segmentation includes 126 labels. Manual refinement of the 17 atlas images required approximately 1–4 hours per atlas depending on gestational age (with later gestational weeks requiring more extensive corrections). Corrections were focused primarily on boundaries between the cortical plate, white matter, and cerebrospinal fluid (CSF), as well as the incorporation of labels new to CRL-2025. Tissue and region atlas segmentations can be crossed to produce cortical parcellations.

\subsection*{Fetal diffusion MRI atlas and labels}
Construction of the fetal dMRI atlas was done previously as described in detail in Khan et al.~\cite{khan2019fetal}, which produced a population-averaged diffusion tensor imaging (DTI) atlas of the fetal brain spanning 23 to 38 gestational week. GAs of 21 and 22 weeks were not included due to insufficient diffusion data. 
Later, Calixto et al.~\cite{calixto2025detailed}, added tissue segmentations and cortical parcellations to the fetal DTI atlas. These labels were initialized on the diffusion atlas through diffeomorphic registration of the corresponding CRL age-matched T2-weighted atlas to the MD map of the DTI atlas. Using the resulting deformation fields, labels from the age-matched fetal brain atlases were propagated to each gestational age of the diffusion atlas. To improve anatomical consistency and reduce local label misalignments, a multi-atlas segmentation fusion strategy was subsequently applied following the approach of Karimi et al.~\cite{karimi2025detailed}. For atlas weeks 23–35, the propagated and fused labels for major tissue classes and anatomically defined cortical and subcortical structures were carefully reviewed and manually refined by an expert with more than five years of experience in fetal neuroanatomy on MRI and dMRI and subsequently verified by a neuroradiologist with more than ten years of experience in fetal MRI~\cite{calixto2025detailed}. The label scheme for the diffusion atlases omits the following structures in both the tissue and regional segmentation: fornix, subthalamic nucleus, hippocampal commissure, and all CSF spaces. For the current work, tissue and regional segmentations were diffeomorphically propagated from these pre-existing segmentations to atlas weeks 36-38, quality controlled, and manually refined for boundary coherency and accuracy.

Finally, to facilitate multimodal use of the CRL-2025 T2-weighted and diffusion atlases, we standardized the diffusion atlas header information and computed rigid transformations from each diffusion atlas age to the corresponding T2-weighted atlas space using the diffusion-derived MD map. The released diffusion atlas preserves its original diffusion grid with corrected spatial metadata, and the rigid transformation matrices are provided for users who wish to map diffusion tensor fields, scalar maps, or label maps to the T2 atlas space. When applying these transformations, tensor images should be resampled with tensor-aware interpolation and reorientation (e.g., \texttt{antsApplyTransforms -e 2}), scalar diffusion-derived maps such as FA and MD can be resampled with B-spline interpolation, and tissue or regional labels should be resampled with multi-label interpolation to preserve discrete label identities.


\subsection*{Automatic Segmentation}
\label{sec_automaticSegmentation}
Many advanced downstream tasks in quantitative fetal brain MRI analysis rely on regional and/or tissue-level segmentations. The CRL-2025 atlas uniquely provides labels on both anatomical and diffusion MRI atlases that enable automatic segmentations. We have explored and developed two mainstream techniques for automatic fetal brain MRI segmentation based on these atlases: 1) multi-atlas segmentation, and 2) deep-learning based segmentation, detailed in the following subsections.

\paragraph{Multi-atlas segmentation}
Multi-atlas segmentation (MAS) involves three main steps: 1) deformable registration of atlases/templates to the query subject anatomy, 2) applying the calculated deformations to propagate labels from the atlases/templates to the query subject space, and 3) label fusion to compute a consensus label for each voxel of the query image. Our MAS pipeline uses ANTs diffeomorphic deformable registration~\cite{Avants2008} with hierarchical rigid, affine, and symmetric normalization by maximizing cross-correlation similarity metric for registration, interpolation for label propagation, and probabilistic STAPLE~\cite{AkhondiAsl2013} for label fusion. However, various other tools, such as MIRTK deformable registration~\cite{rueckert1999nonrigid,schnabel2001generic,schuh2014construction}, other similarity metrics~\cite{avants2011reproducible}, other label fusion techniques, e.g.~\cite{Warfield2004,wang2012multi}, or shape-guided MAS techniques, e.g.~\cite{gholipour2012multi}, can be used. MAS segmentation of reconstructed T2-weighted or DTI fetal brain images should employ age-matched atlases/templates. When using the CRL-2025, for example, one shall use atlases within one week of GA from the query subject's GA. It should be noted that if a larger number of individual-subject templates with reliable labels are available, they can be used as multiple atlases in MAS (separately or in combination with the CRL-2025 atlases). While as few as three atlas images may be sufficient for major structures, MAS literature shows that using larger number of atlases generally improves MAS accuracy~\cite{iglesias2015multi} with quality and diversity of atlases also serving as important factors. For more detailed description of MAS and its performance we refer to the literature cited above, and for an application and validation in fetal MRI we refer to Gholipour et al.~\cite{gholipour2017normative}.

\paragraph{Deep learning segmentation}

\indent 
Deep learning based segmentation models have gained popularity due to their fast inference capabilities compared to conventional MAS methods, as well as their ability to learn direct mappings from images to segmentations without requiring image registration or label propagation. We developed, trained and evaluated several deep learning models for fetal brain MRI segmentation. Specifically, we trained and evaluated four segmentation models based on state-of-the-art network architectures, in specific 1) nnU-Net~\cite{isensee2021nnu}, 2) UNETR~\cite{hatamizadeh2021unetr}, 3) SwinUNETR~\cite{hatamizadeh2021swin} and 4) a network with Residual Vision MAMBA (RVM) blocks in a U-Net backbone with depthwise separable convolutions (DWConvs)~\cite{liao2024lightmunet}.
Among these models, the model based on nnU-Net, which is based on the 3D Dynamic U-Net architecture implemented in MONAI~\cite{cardoso2022monai}, performed best. This model, referred here as FetalSEG, operates on input patches of size 128×128×128 voxels. The architecture consists of six resolution levels with five downsampling stages. Each stage uses 3×3×3 convolution kernels and stride-2 convolutions for downsampling. Instance normalization and LeakyReLU activations are used after each convolution. The decoder mirrors the encoder using transposed convolutions with corresponding strides for upsampling. Residual blocks and deep supervision were disabled in our configuration.

Models were implemented in PyTorch. All experiments, including training and testing, were performed on a computer workstation with an NVIDIA RTX A6000 GPU for 200 epochs using the AdamW optimizer~\cite{loshchilov2017decoupled} with a learning rate of $10^{-4}$ and a PolyLR scheduler with a batch size of one. We trained, tested, and compared deep learning models on an in-house dataset of reconstructed T2-weighted fetal MRI scans of 177 fetuses scanned one or multiple times in the GA range of 20–38 weeks. The dataset was split subject-wise into 106 training, 45 testing, and 26 validation samples. Tissue segmentations on these images were generated as part of previous projects through manual refinement of multi-atlas segmentations following the manual segmentation protocols described in the Methods section based on CRL-2017~\cite{gholipour2017normative}. During training, data augmentation was applied to improve model robustness and generalization. Random spatial transformations (including flipping, rotation, zooming, and affine transformations) were applied consistently to both input images and corresponding segmentation labels, while intensity-based augmentations (Gaussian noise, bias field perturbations, and smoothing) were applied to the input images.
$
$

\section*{Data Records}
\label{sec_datarecords}

As part of this data descriptor, we release several related but distinct resources. First, the CRL-2025 T2-weighted structural atlas was newly constructed from 193 reconstructed fetal T2-weighted MRI scans from 159 fetuses and spans 21--37 gestational weeks. Second, the processed subject-level T2-weighted reconstructions used as input for atlas construction are released to provide input--output transparency and enable reproducibility. Third, the CRL fetal diffusion MRI atlas, based on motion-robust diffusion tensor image (DTI) reconstruction, originally constructed by Khan et al.~\cite{khan2019fetal}, is re-released here with corrected header information, rigid transformation matrices to the corresponding CRL-2025 T2-weighted atlas space, and with tissue and regional labels. Fourth, the T2-weighted and diffusion atlas label sets are provided as separate label-key files, reflecting their modality-specific labeling schemes. Finally, FetalSEG is released as an auxiliary deep-learning tool for automatic segmentation of individual reconstructed T2-weighted fetal MRI scans; it was trained on individual subject images labeled using the CRL-2017 scheme and was not used for CRL-2025 atlas construction. The differences between the CRL-2017 and CRL-2025 labeling schemes are shown in Table XX. Together, these resources provide complementary structural, diffusion, labeling, and segmentation tools for fetal brain MRI analysis.

\paragraph{CRL-2025 spatiotemporal T2 and diffusion atlases}
The CRL-2025 fetal T2 brain MRI atlas and the re-release of the CRL fetal diffusion brain MRI atlas~\cite{khan2019fetal} are publicly available as Harvard Dataverse datasets \url{https://dataverse.harvard.edu/dataverse/CRL2025Atlas/}, and via the DOI links \url{https://doi.org/10.7910/DVN/QOO75G} (T2w)\cite{DVN/QOO75G_2025} and \url{https://doi.org/10.7910/DVN/XWKCIE} (DTI)\cite{DVN/XWKCIE_2025}. The T2 atlas spans gestational ages (GA) 21 to 37 while the diffusion atlas spans GA weeks 23 to 38. Each atlas includes tissue segmentations and anatomical parcellations (regional). Datasets are provided as NIfTI files (.nii.gz) organized by gestational week and modality. Three accompanying label-key text files are included: \texttt{T2WAtlas\_labelkey-tissue.txt}, for the T2w tissue segmentations; \texttt{T2WAtlas\_labelkey-region.txt}, for the T2w regional parcellations; and \texttt{DWIAtlas-labelkey.txt}, which corresponds to a modified version of the T2w labeling schemes used in the DTI atlases. All files can be viewed using standard neuroimaging software such as ITK-SNAP~\cite{yushkevich2006user}.

\paragraph{Subject-level dataset}
We have released the full set of reconstructed fetal MRI datasets used as inputs to the CRL-2025 T2 atlas, publicly available at \url{https://dataverse.harvard.edu/dataverse/CRL2025Atlas/} (DOI: \url{https://doi.org/10.7910/DVN/A7YSH7})\cite{DVN/A7YSH7_2025}, titled "CRL2025 Fetal Dataset".
These primary data are published here for the first time and include all reconstructed subject-level T2-weighted (T2W) volumes and associated metadata (gestational age, reconstruction method, and atlas gestational week categorizations). All images were reconstructed from multiple T2-weighted stacks, intensity corrected and registered to atlas space. The original MRI acquisitions were obtained under multiple IRB-approved studies at Boston Children’s Hospital, and all data were fully de-identified prior to sharing in accordance with institutional IRB protocols and Scientific Data’s Human Data requirements. No new human data were collected specifically for this study; instead, previously acquired research scans were reconstructed, processed, and curated by the authors to create the subject-level dataset.
README files included in each dataset folder describe file organization and naming conventions. The trained models can be found at \url{https://dataverse.harvard.edu/dataverse/CRL2025Atlas/}  (DOI: {\url{https://doi.org/10.7910/DVN/A98K4K}})\cite{DVN/A98K4K_2026}.

\section*{Technical Validation}
\label{sec_validation}

\subsection*{The spatiotemporal fetal brain MRI atlas}
\label{sec_atlas}

\begin{figure}[!h]
\centering
\includegraphics[width=15cm]{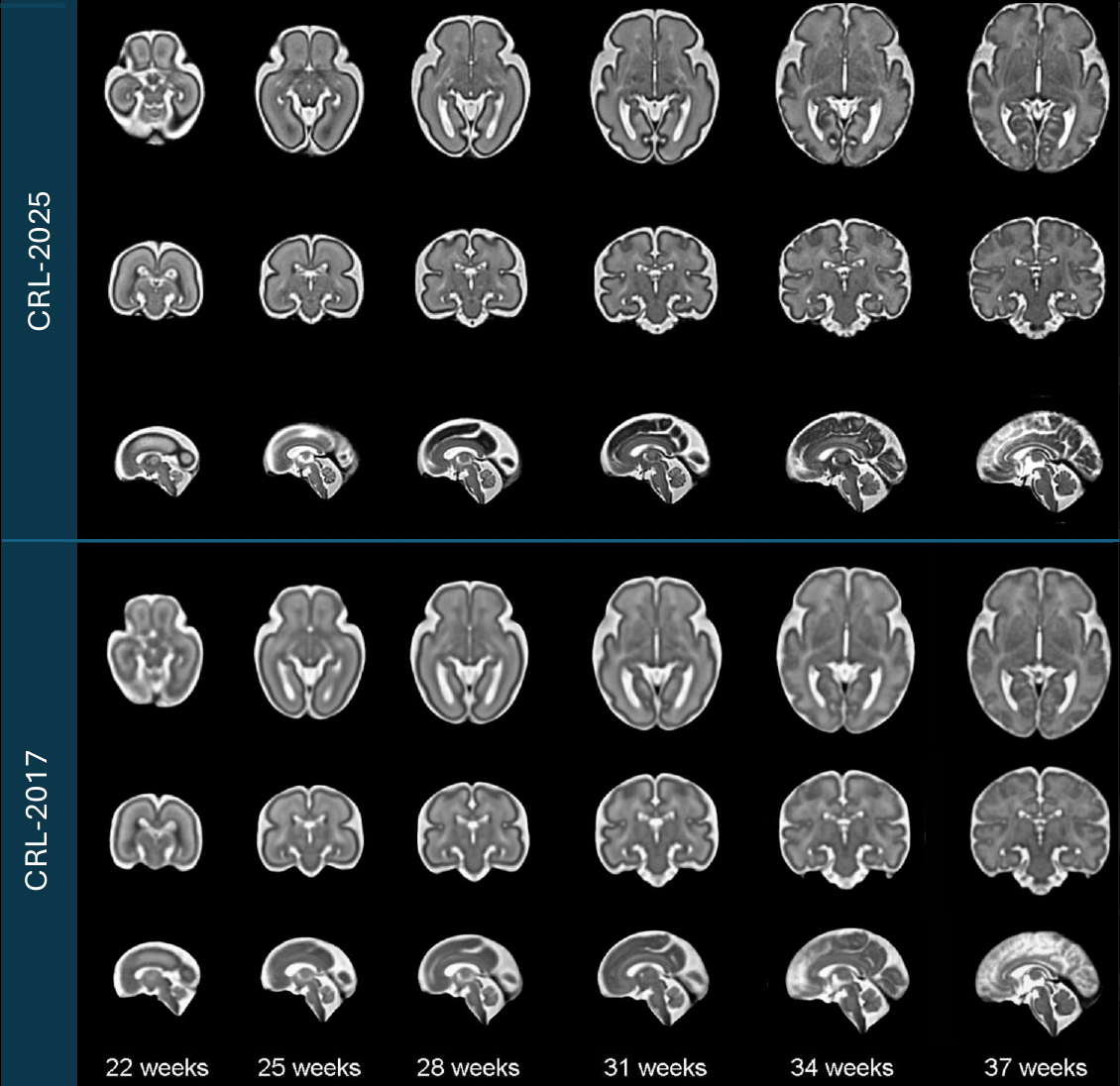}
\caption{Comparison of spatiotemporal fetal brain MRI atlases ( CRL-2025 vs. CRL-2017) at six representative gestational ages: 22, 25, 28, 31, 34, and 37 weeks. Axial, coronal, and sagittal views are presented for each atlas at each age.}
\label{fig_atlas}
\end{figure}

Fetal brain MRI scans were pre-processed following the procedures outlined in the Methods section. The processed images were then used to construct the spatiotemporal atlas according to Algorithm \ref{alg:atlas_construction}, which enables the generation of an unbiased average atlas at any continuous gestational age. Figure \ref{fig_atlas} displays axial, coronal, and sagittal views of the resulting atlas at representative gestational ages, alongside corresponding views from the CRL-2017 atlas for comparison~\cite{gholipour2017normative}. Both visual inspection and quantitative analyses suggest that the CRL-2025 atlas much better preserves anatomical details and is sharper than the CRL-2017 atlas. For this, as shown in Table \ref{tab:median_sharpness_2017_2025}, we computed median edge sharpness to quantitatively assess the clarity of anatomical boundaries in each atlas. This metric, defined as the median gradient magnitude at anatomical edges detected within the atlas, reflects sharpness of typical tissue transitions. Across all gestational ages, CRL-2025 consistently exhibited higher median edge sharpness than CRL-2017, with differences ranging from +0.14 to +0.39. These results demonstrate that CRL-2025 provides more clearly defined anatomical boundaries throughout development, confirming its advantage for many applications such as atlas-based segmentation and analysis.
Also, the atlas shape and size complies with the dHCP spatio-temporal fetal brain MRI atlas.
\begin{table}[h!]
\centering
\caption{Median edge sharpness (gradient magnitude at anatomical boundaries) 
for CRL-2017 and CRL-2025 atlases across gestational ages, shows superior delineation of the anatomy in CRL-2025 compared to CRL-2017.}
\label{tab:median_sharpness_2017_2025}
\begin{tabular}{@{}lcccccccc@{}}
\toprule
\textbf{GA} & 22 & 23 & 24 & 25 & 26 & 27 & 28 & 29 \\
\midrule
CRL-2017 & 1.008 & 0.938 & 0.986 & 0.950 & 0.911 & 0.884 & 0.886 & 0.883 \\
CRL-2025 & 1.349 & 1.330 & 1.290 & 1.270 & 1.266 & 1.229 & 1.225 & 1.160 \\
\midrule
\textbf{GA} & 30 & 31 & 32 & 33 & 34 & 35 & 36 & 37 \\
\midrule
CRL-2017 & 0.867 & 0.842 & 0.809 & 0.760 & 0.696 & 0.632 & 0.605 & 0.622 \\
CRL-2025 & 1.159 & 1.177 & 1.093 & 1.037 & 0.963 & 0.848 & 0.799 & 0.764 \\
\bottomrule
\end{tabular}
\end{table}

\begin{figure}[!h]
\centering
\includegraphics[width=17cm]{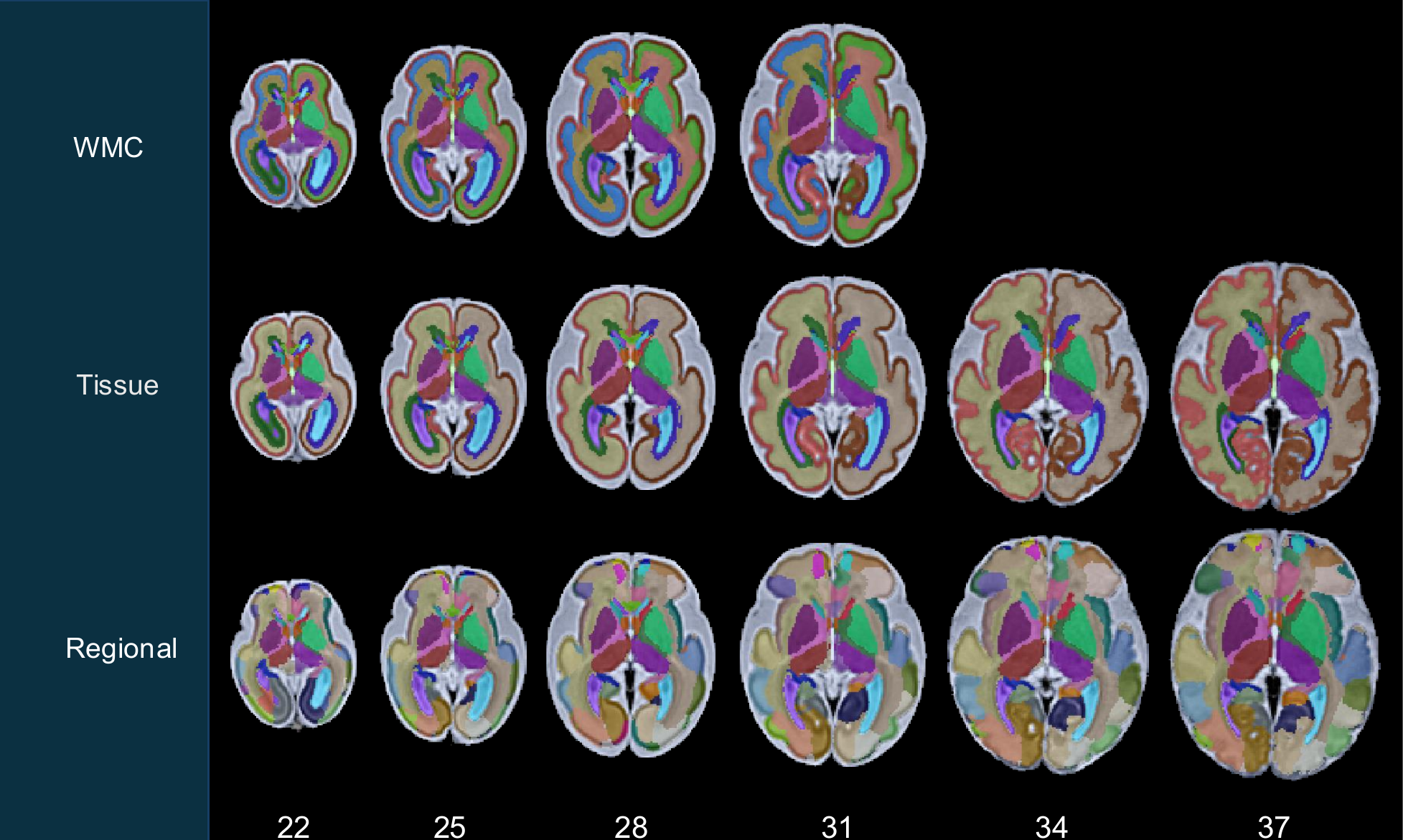}
\caption{Tissue and regional segmentations and structural labels overlaid on axial views of the CRL-2025 spatiotemporal fetal brain MRI
atlas at six representative gestational age (GA) weeks. All label schemes have subcortical structures including lentiform and caudate nuclei, internal capsules, thalami, and hippocampi separately on each hemisphere. Tissue segmentation labels (middle row) delineate the cortical plate-white matter boundary, CSF, and subcortical structures. In the top row, the white matter of the tissue segmentation is divided into white matter compartments (WMC) including the ventricular and intermediate zones and the subplate. These transient WMCs gradually disappear and were not clearly observable on the atlases beyond 31 weeks. The bottom row displays regional segmentations which are useful for regional and connectivity analyses.}
\label{fig_atlaslabels}
\end{figure}

\begin{figure}[!h]
\centering
\includegraphics[width=17cm]{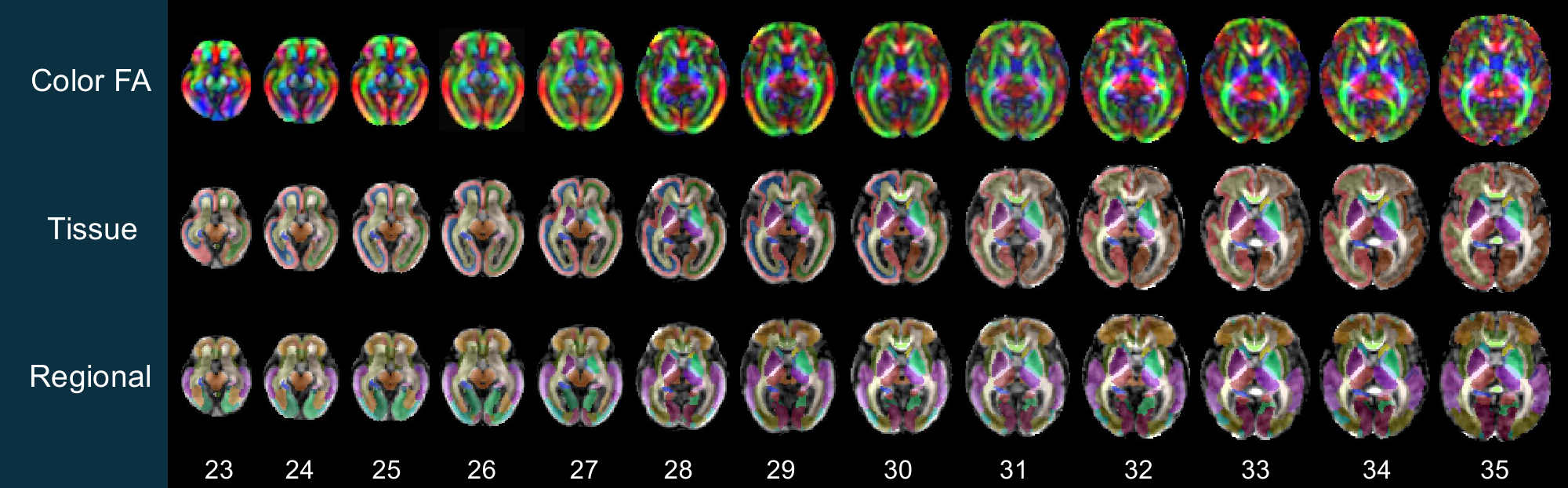}
\caption{The spatiotemporal fetal brain diffusion MRI atlas shown at representative GA weeks 23-35. Top row: Color fractional anisotropy (FA). Middle row: Tissue segmentation. Bottom row: Regional segmentation.}
\label{fig_dwiatlaslabels}
\end{figure}

\indent Labels were generated on the spatiotemporal fetal brain MRI atlases following the procedure described in the Methods section. As illustrated in Figure \ref{fig_atlaslabels}, these labels include 1) the transient compartments of the developing white matter (WMC) for all atlases less than 32 weeks, 2) tissue types, and 3) regions. For details of these labels, we refer to the text files in the data repository. Figure \ref{fig_dwiatlaslabels} shows the fetal DTI atlas~\cite{khan2019fetal} (color anisotropy in the first row) along with its tissue segmentations (second row) and regional parcellations (third row) that are released in this edition with the CRL-2025 atlas (GA 36 and 37 are not shown in this figure). All of the segmentations and labels on the T2w atlas were carefully checked, manually refined, and validated in multiple rounds by four experts under the supervision of a neurologist and a neuroanatomist each with more than a decade of experience in assessing fetal neuroanatomy on MRI. All the segmentations on the DTI atlas were carefully checked, manually corrected, and validated by two experts with several years of experience in fetal neuroanatomy on MRI and dMRI. To further assess and validate the use of the atlases and their labels, we used the atlases for multi-atlas segmentation of individual subject fetal brain MRIs, and used multi-atlas generated, manually-corrected labels on individual subject images~\cite{gholipour2017normative} to train and test deep learning based segmentation models (as discussed in the Methods section) and presented in the next section.



%

\subsection*{Automatic Segmentation}
\label{auto_segmentation}

\begin{figure}[!h]
\centering
\includegraphics[width=1\linewidth]{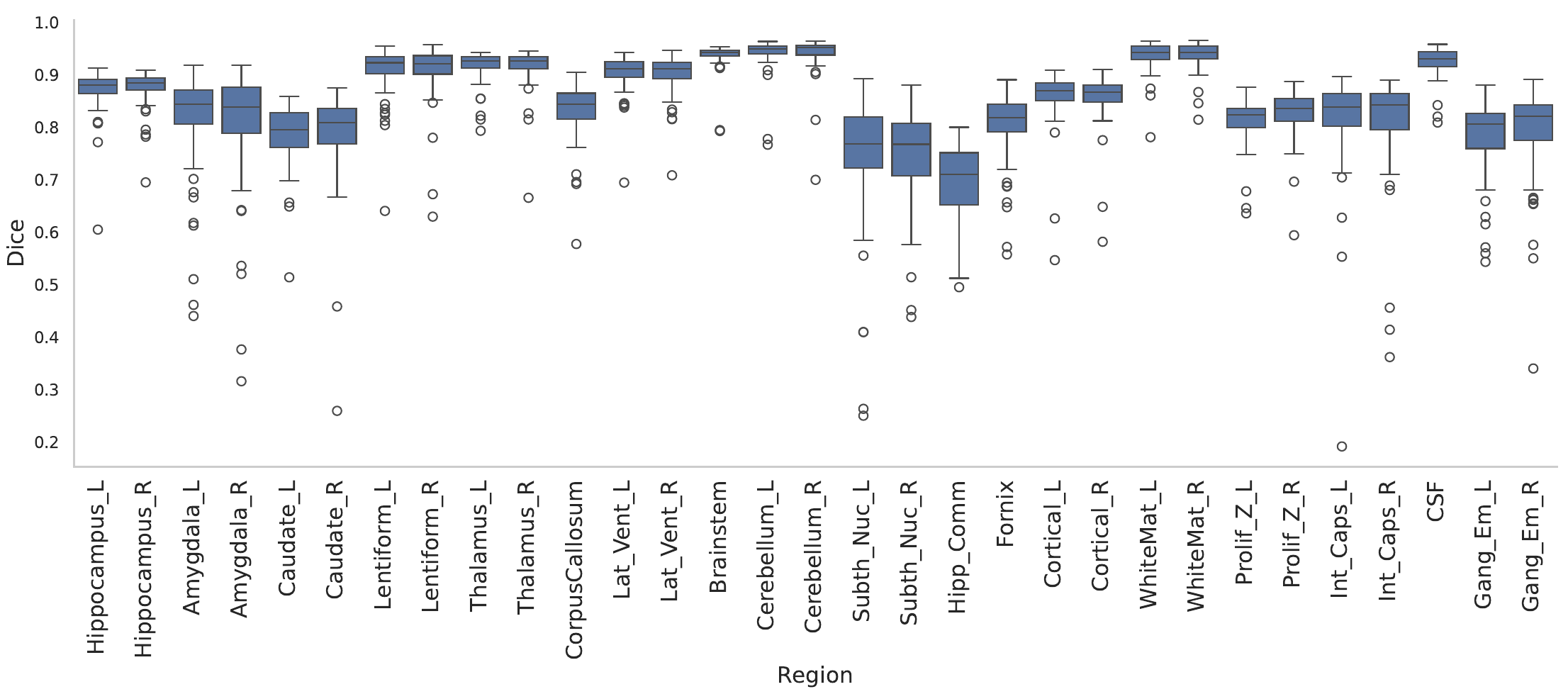}
\vspace{1em} 
\caption{Boxplots of Dice performance metric over the test sets for 31 distinct segmentation regions using the automatic segmentation model, FetalSEG.}
\label{fig:BCH}
\end{figure}

\indent Table~\ref{tab:results} summarizes the automatic segmentation performance of FetalSEG. 
Fig.~\ref{fig:BCH} presents boxplots of the DSC across the 31 segmentation regions for the test subjects. The results indicate that FetalSEG achieved reliable automatic segmentations for major structures such as the cortical plate, subplate, CSF, ventricles, corpus callosum, and thalamus, whereas small or low-contrast structures such as the amygdala, caudate nuclei, subthalamic nuclei, hippocampal commissure, and fornix remained challenging.


\begin{table}[!h]
\centering
\caption{The performance of FetalSEG for automatic fetal brain MRI tissue segmentation: average and standard deviation of Dice and HD values}
\label{tab:results}

\begin{tabular}{lcc}
\toprule
 & \multicolumn{2}{c}{ Over 31 regions} \\
\cmidrule(lr){2-3}
Model & Dice & HD \\
\midrule
FetalSEG & \textbf{0.853 $\pm$ 0.09} & \textbf{0.984 $\pm$ 0.44} \\
\bottomrule
\end{tabular}
\end{table}

\newpage
\section*{Usage Notes}
\label{sec_usage}
The CRL-2025 fetal brain MRI atlases are provided in standard NIfTI format with accompanying label key text files for use in common neuroimaging software such as ITK-SNAP, FSL, and ANTs. Users are advised to use gestational-age-matched atlas versions for optimal accuracy in atlas-based segmentation. Example workflows and scripts for data visualization, preprocessing, and segmentation are available in the linked code repositories. The atlas and the related tools can be used for spatial normalization, registration, segmentation, and parcellation for group level or individual level image analysis studies. The atlases are not intended for use in clinical diagnosis, prognosis, or evaluation, and may not be used in making inferences in any clinical practice including but not limited to any measurements that may affect clinical care or clinical decision making.

\section*{Data Availability}
All data supporting this study, including the CRL-2025 dataset and spatiotemporal fetal brain MRI atlases, tissue segmentations, and diffusion MRI atlases, are publicly available through the Harvard Dataverse repository: \url{https://dataverse.harvard.edu/dataverse/CRL2025Atlas}
.
\section*{Code Availability}

The multi-atlas segmentation code can be found at \url{https://github.com/IntelligentImaging/CRL-2025-MAS}. The deep learning based segmentation codes can be found at \url{https://github.com/IntelligentImaging/CRL-2025}.
These repositories include resources for customizing workflows for automatic segmentation model training and deployment, links to the trained models, and data analysis tools to reproduce figures presented in the manuscript. The trained models can be found at \url{https://dataverse.harvard.edu/dataverse/CRL2025Atlas/},  DOI: {\url{https://doi.org/10.7910/DVN/A98K4K}}
\label{sec_codeAvailability}.

\section*{Author Contributions}
M.B. developed and validated image segmentation methods, conducted experiments, processed data and analyzed results, and wrote the article. C.V.A. developed and validated atlas construction, atlas-based segmentation, and atlas labeling methods, developed image processing pipelines, acquired data, processed and analyzed data, and wrote the article. J.W. developed and validated atlas construction methods, and wrote the article. R.F. developed and validated image segmentation methods, processed data and analyzed results, and wrote the article. S.K. developed and validated atlas construction methods, processed data and analyzed results, and reviewed the article. C.C. developed and validated atlas labeling, processed data and analyzed results, and reviewed the article. C.J. developed and validated atlas labeling, processed data and analyzed results, and reviewed the article. L.V. developed and validated atlas labeling, processed data and analyzed results, and reviewed the article. A.O. developed and validated atlas labeling, processed data and analyzed results, and reviewed the article. O.A. developed and evaluated imaging methods, acquired and analyzed data, and reviewed the article. S.K.W. developed imaging and image processing methods, analyzed results, funded and supervised the project, and reviewed the article. C.K.R. developed and validated atlas labeling, processed data and analyzed results, funded and supervised the project, and wrote the article. A.G. developed and validated atlas construction, image segmentation, and atlas labeling methods, acquired data, analyzed results, funded and supervised the project, and wrote the article. All authors reviewed the manuscript.

\section*{Competing Interests}
The authors declare no competing interests.

\section*{Acknowledgements}
This study was supported in part by the National Institutes of Health (NIH) through grants R01EB013248, R01EB018988, R01NS106030, R01NS121334, R01EB031849, R01EB032366, R01HD109395, and K23NS101120; in part by the Office of the Director of the NIH under award number S10OD025111; in part by the National Science Foundation under grant number 212306; and in part by the Thrasher Research Fund, the Fetal Health Foundation, and the McKnight Foundation. This research was also partly supported by NVIDIA Corporation and utilized NVIDIA RTX A6000 GPUs. The content of this publication is solely the responsibility of the authors and does not necessarily represent the official views of the NIH, NSF, or NVIDIA.

The authors acknowledge the use of the MIRTK software toolkit for image registration (\url{https://mirtk.github.io/}), as well as ANTS (\url{https://stnava.github.io/ANTs/}), N4ITK, and ITKSNAP (\url{https://www.itksnap.org/}) developed in the Penn Image Computing and Science Laboratory (PICSL) (\url{http://picsl.upenn.edu/software/}) at the University of Pennsylvania.
\bibliography{Gholipour_Fetal_Brain_Atlas}

\end{document}